# Convection in a volatile nitrogen-ice-rich layer drives Pluto's geological vigor

William B. McKinnon[1], Francis Nimmo[2], Teresa Wong[1], Paul M. Schenk[3], Oliver L. White[4], J. H. Roberts[5], J. M. Moore[4], J. R. Spencer[6], A. D. Howard[7], O. M. Umurhan[4], S. A. Stern[6], H. A. Weaver[5], C.B. Olkin[6], L. A. Young[6], K. E. Smith[4] and the New Horizons Geology, Geophysics and Imaging Theme Team; [1]Deptartment of Earth and Planetary Sciences and McDonnell Center for the Space Sciences, Washington University in St Louis, St. Louis MO, 63130, USA; [2]Deptartment of Earth and Planetary Sciences, University of California Santa Cruz, Santa Cruz CA, 95064, USA; [3]Lunar and Planetary Institute, Houston TX, 77058, USA; [4]National Aeronautics and Space Administration (NASA) Ames Research Center, Moffett Field, CA, 94035, USA; [5]Johns Hopkins University Applied Physics Laboratory, Laurel MD, 20723, USA; [6]Southwest Research Institute, Boulder CO, 80302, USA; [7]Department of Environmental Sciences, University of Virginia, Charlottesville VA 22904

**The vast, deep, volatile-ice-filled basin informally named Sputnik Planum is central to Pluto's geological activity[1,2]. Composed of molecular nitrogen, methane, and carbon monoxide ices[3], but dominated by $N_2$-ice, this ice layer is organized into cells or polygons, typically ~10-40 km across, that resemble the surface manifestation of solid state convection[1,2]. Here we report, based on available rheological measurements[4], that solid layers of $N_2$ ice $\gtrsim$1 km thick should convect for estimated present-day heat flow conditions on Pluto. More importantly, we show numerically**

**that convective overturn in a several-km-thick layer of solid nitrogen can explain the great lateral width of the cells. The temperature dependence of $N_2$-ice viscosity implies that the SP ice layer convects in the so-called sluggish lid regime[5], a unique convective mode heretofore not definitively observed in the Solar System. Average surface horizontal velocities of a few cm/yr imply surface transport or renewal times of ~500,000 years, well under the 10 Myr upper limit crater retention age for Sputnik Planum[2]. Similar convective surface renewal may also occur on other dwarf planets in the Kuiper belt, which may help explain the high albedos of some of them.**

Sputnik Planum (SP) is the most prominent geological feature on Pluto revealed by NASA's New Horizons mission. It is a ~900,000 km$^2$ oval-shaped unit of high-albedo plains (Fig. 1a) set within a topographic basin at least 2-3 km deep (Fig. 1b). The basin's scale, depth and ellipticity (~1300 x 1000 km), and rugged surrounding mountains, suggest an origin as a huge impact – one of similar scale to its parent body as Hellas on Mars or South Pole-Aitken on the Moon[6]. The central and northern regions of SP display a distinct cellular/polygonal pattern (Fig. 1c). In the bright central portion, the cells are bounded by shallow troughs locally up to 100 m deep (Fig. 1d), and the centers of at least some cells are elevated by ~50 m relative to their edges[2]. The southern region and eastern margin of SP do not display cellular morphology, but instead show featureless plains and dense concentrations of km-scale pits[2].

No impact craters have been confirmed on SP in New Horizons mapping at 350 m/pixel scale, nor in high-resolution strips (resolutions as fine as 80 m/pixel). The crater retention age of SP is very young, no more than ~10 Myr based on models of the impact flux of small Kuiper belt objects onto Pluto[7]. This indicates renewal, burial, or erosion of the surface on this time scale or shorter. Evidence for all three processes is seen in the form of possible convective overturn, glacial inflow of volatile ice from higher standing terrains at the eastern margin, and likely sublimation landforms such as the pits[2]. In addition, the apparent flow lines around obstacles in northern SP and the pronounced distortion of some fields of pits in southern SP are evidence for the lateral, advective flow of SP ices[1,2].

From New Horizons spectroscopic mapping, $N_2$, $CH_4$ and CO ice all concentrate within Sputnik Planum[3]. All three ices are weak, van der Waals bonded molecular solids and are not expected to be able to support appreciable surface topography over any great length of geologic time[4,8–10], even at the present surface ice temperature of Pluto (37 K )[1]. This is consistent with the overall smoothness of SP over 100s of kilometers (Fig. 1b). Convective overturn that reaches the surface would also eliminate impact and other features, and below we numerically estimate the time scale for SP's surface renewal.

Quantitative radiative transfer modeling of the relative surface abundances of $N_2$, $CH_4$, and CO ices within SP[11] shows that $N_2$ ice dominates $CH_4$ ice within SP, especially in the central portion of the planum (the bright cellular plains) where the cellular structure is best defined topographically (Fig. 1d). Ices at depth need not match the surface composition, but continuous exposure (such as by convection) makes this more likely. $N_2$ and CO ice have nearly the same density (≈1.0 g/cm$^3$), whereas $CH_4$ ice is half as dense[2]. Hence water ice blocks can float in solid $N_2$ or CO, but not in solid $CH_4$. Water ice has been identified in the rugged mountains that surround SP[3], and blocks and other debris shed from the mountains at SP's periphery appear to be floating[2]; moreover, glacial inflow appears to carry along water-ice blocks, and these blocks almost exclusively congregate at the margins of the cells/polygons, consistent with being dragged to the downwelling limbs of convective cells (Fig. 2a). This indicates that while $CH_4$ ice is present within SP, it is not likely to be volumetrically dominant. In terms of convection we concentrate on the rheology of $N_2$ ice.

Deformation experiments for $N_2$ ice show mild power-law creep behavior (strain rate proportional to stress to the $n = 2.2 \pm 0.2$ power) and a modest temperature dependence of its viscosity[4]. $N_2$ Diffusion creep ($n$ =1) has also been predicted[10,12], but not yet observed experimentally. Convection in a layer occurs if the critical Rayleigh number ($Ra_{cr}$) is exceeded. The Rayleigh number, the dimensionless measure of the vigor of convection, for a power-law fluid heated from below is given by[13]

$$Ra = \frac{\rho g \alpha A^{1/n} \Delta T D^{(2+n)/n}}{\kappa^{1/n} \exp(E^*/nRT)} , \quad (1)$$

where $D$ is the thickness of the convecting layer, $\kappa$ is the thermal diffusivity, $g$ is gravity, $\rho$ the ice layer density, $\alpha$ the volume thermal expansivity, $\Delta T$ the superadiabatic temperature drop across the layer, and $A$ is the preexponential constant in the stress strain-rate relationship, $E^*$ is the activation energy of the dominant creep mechanism, and $R$ is the gas constant.

The critical Rayleigh number depends on the temperature drop and the associated change in viscosity[13], as deformation mechanisms are activated processes. For a given $\Delta T$, the $Ra_{cr}$ implies a critical or minimum layer thickness, $D_{cr}$, below which convection cannot occur. This is illustrated in Fig. 3 for $N_2$ ice. We assume an average ice surface temperature of 36 K set by vapor-pressure equilibrium over an orbital cycle[14], and an upper limit on the basal temperature set by the $N_2$ ice melting temperature of 63 K (ref. 15). From Fig. 3 we conclude that convection in solid nitrogen on Pluto is a facile process: critical thicknesses are generally low, less than 1 kilometer, as long as the necessary temperatures at depth are achieved.

The temperature profile in the absence of convection is determined by conduction. $N_2$ ice has a low thermal conductivity[15], which together with a present-day radiogenic heat flux for Pluto of roughly 3 mW $m^{-2}$ implies a conductive temperature gradient of ~15 K $km^{-1}$. Over Pluto's history, radiogenic heat has dominated Pluto's internal energy budget[16,17]; we argue that relatively unfractionated, solar composition carbonaceous chondrite is the best model for the rock component of worlds accreted in the cold, distant regions of the Solar System[16]. The abundances of U, Th, and $^{40}K$ are consistent across the most primitive individual examples of this meteorite clan (the CI chondrites), to within 15% (ref. 18), and Pluto's density implies that ≈2/3 of its mass could be composed of solar composition rock (the rest being ices and carbonaceous material)[19]. Nevertheless, regional and temporal variations in heat flow are possible, so Fig. 3 illustrates the temperatures reached as a function of depth, with the conclusion being that even under broad variations in heat flow, temperatures sufficient to drive convection in SP are plausible for $N_2$-ice layers thicker than ~500 meters.

Clearly, the horizontal scale of the cells in SP (Figs. 1d, 2a,b) should reflect the vertical scale (depth) of the SP basin ice fill, but this presents a problem. For isoviscous Rayleigh-Bérnard convection, the aspect ratio (width/depth) of well-developed convection cells is near unity. Numerical calculations by us for Newtonian and non-Newtonian convection in very wide 2D domains, but without temperature-dependent viscosity, give aspect ratios near 1 (Methods). If the cells/polygons on Sputnik Planum are the surface expression of convective cells, then 20-to-40–km cell diameters (wavelengths $\lambda$) imply depths to the base of the $N_2$ ice layer in SP of about 10-to-20 km. This is very deep, and much deeper than any likely impact basin, especially as the surface of SP is already at least 2-3 km deep compared with the rest of Pluto (Fig 1b). The deepest impact basins of comparable scale known on any major icy world are on Iapetus, a body of much lower density (hence lower rock abundance and heat flow) and surface gravity than Pluto. Gravity scaling from basins on Iapetus[20], we estimate the SP basin was initially no deeper than ~10 kilometers total (that is, before filling by volatile ices or any isostatic adjustment).

The solution to this apparent problem is likely the temperature dependence of the $N_2$ ice viscosity. Given that the maximum $\Delta T$ across the SP $N_2$ layer is 27 K, the maximum corresponding Arrhenius viscosity ratio ($\Delta\eta$) for the experimentally constrained activation energy is ~150 (Methods); if we adopt the (larger) activation energy for volume diffusion[21], this ratio potentially increases to ~2 x $10^5$. This potential range in $\Delta\eta$ strongly suggests that SP convects in the sluggish lid regime[5,13,22]. In sluggish lid convection the surface is in motion and transports heat, but moves at a much slower pace than the deeper, warmer subsurface. A defining characteristic of the sluggish lid regime — depending on $Ra_b$ (the Rayleigh number defined with the basal viscosity) and $\Delta\eta$ — are large aspect ratio convection cells. This differs from isoviscous convection in which the aspect ratios are closer to one, or at the other end of the viscosity contrast spectrum, stagnant lid convection, in which aspect ratios are again closer to one but confined (“hidden”) beneath an immobile, high-viscosity surface layer.

We illustrate such temperature-dependent viscosity convection numerically, using the finite element code CitCom[22] (a typical example is shown in Fig. 4). Given that $N_2$- ice rheology is imprecisely known (unlike well-studied geological materials such as olivine or water ice), we survey different combinations of $Ra_b$ and $\Delta\eta$ in a Newtonian framework (similar to previous work[5,22]), but with a rigid (no-slip) lower boundary condition appropriate to the SP ice layer (Methods). We find that aspect ratios easily reach values of 2 or 3 (or $\lambda/D$ of 4 or 6), regardless of initial perturbation wavelength. In this case cell dimensions between 20 and 40 km across could imply a layer thicknesses as small as ~3 to 6 kilometers. We note that while these depths are not excessive, they are deep enough to carry buoyant, km-scale water ice blocks. In addition, simulations with a free-slip lower boundary, which would apply to SP ice that is at or near melting at its base, yield aspect ratios as great as ~6 ($\lambda/D$ ~12).

Numerical simulations can be tested with SP by assuming reasonable heat flows (say, chondritic within ±50%) and comparing the resulting dynamic topography with that observed. Non-dimensional surface horizontal velocity $\upsilon_x$, normal stress $\sigma_{zz}$, and heat flow $q_z$ for the example calculation are shown in Figs. 4b-c. To dimensionalize we choose $D$ = 4.5 km and $\Delta T$ = 20 K to match the typical horizontal scale of the cells (e.g., nearly 30 km, with a convective aspect ratio of 3) and give a chondritic heat flow (see Methods). The dynamic topography due to the thermal buoyancy of the flow is given by $\sigma_{zz} /\rho g$, and its scale is given at the right hand side of Fig. 4c. This dynamic topography is consistent with available measurements[1,2]. Average surface velocities (Fig. 4b) in this example are a few cm/yr, which for the horizontal scale of cells on SP translates into a time scale to transport surface ice from the center of a given upwelling to the downwelling perimeter of ~500,000 years. This is well within the upper limit for the crater retention age for the planum, ~10 Myr (ref. 2). The surface heat flow variation is also notable, nearly double the mean over upwellings and close to zero over downwellings. This means fine scale topography such as pitting or suncups driven by $N_2$ sublimation[2] will be much more stable towards cell/polygonal edges, as the $N_2$ ice there will be as cold and viscous as the surface to considerable depth, which is consistent with the observations of surface texture[2] (e.g., Fig. 2b). We also find slight topographic dimples over downwellings in some of our calculations, which may be related to trough formation at cell edges (Fig. 2b). The troughs themselves, however, are likely to be finite amplitude topographic instabilities of the sort seen on icy satellites elsewhere[23], and are not captured by these convection calculations given that velocities normal to domain boundaries are set to zero.

Convection in a kilometers-thick $N_2$ layer within Pluto's Sputnik Planum basin thus emerges as a compelling explanation for the remarkable appearance of the planum surface (Fig. 1). Sputnik Planum covers 5% of Pluto's surface, so having an $N_2$ ice layer several kilometers deep is equivalent to a global layer of ~200–300 m thick. This is consistent with Pluto's possible total cosmochemical nitrogen inventory[24], especially as Pluto's atmospheric nitrogen escape rate is much lower than previously estimated[25]. For Pluto, SP acts an enormous glacial catchment or drainage basin, the major topographic trap for Pluto's surficial, flowing $N_2$ ice. SP is essentially a vast, frozen sea, one in which convective turnover (now, and even more vigorously in the past) continually refreshes the surface volatile ice inventory. A sealing, superficial lag of less volatile ices or darker tholins cannot develop[24], and the atmospheric cycle of volatile transport is maintained. Moreover, larger Kuiper belt objects (KBOs) are known to be systematically brighter (more reflective) than their smaller KBO cousins[26]. Convective renewal of volatile ice surfaces, as in a basin or basins similar to SP, may be one way in which the small planets of the Kuiper belt maintain their youthful appearance.

**Acknowledgements** New Horizons was built and operated by the Johns Hopkins Applied Physics Laboratory (APL) in Laurel, Maryland, USA, for NASA. We thank the many engineers who have contributed to the success of the New Horizons mission and NASA's Deep Space Network (DSN) for a decade of excellent support to New Horizons. This work was supported by NASA's New Horizons project.

**Author Contributions** W.B.M. led the study and wrote the paper, with significant input from F.N.; T.W. and J.H.R. performed the CitCom finite element convection calculations; P.M.S. developed the software to create stereographic and photoclinometric digital elevation models (DEMs) using New Horizons LORRI and MVIC images, and created the preliminary DEM for Sputnik Planum; O.L.W. mapped the Sputnik Planum region using New Horizons images in ArcGIS; J.M.M., J.R.S., A.D.H, O.M.U. and S.A.S. contributed to the understanding of the multiple roles $N_2$ ice plays in the geology of Sputnik Planum and environs. S.A.S., H.A.W., C.O., L.A.Y. and K.E.S. are the lead scientists of the New Horizons project.

**Author Information** Reprints and permissions information is available at www.nature.com/reprints. The authors declare no competing financial interests. Readers are welcome to comment on the online version of the paper. Correspondence and requests for materials should be addressed to W.B.M. (mckinnon@wustl.edu).

**Methods**

**Mapping and Topography.** The LORRI basemap in Fig 1a was created from the 5 x 4 mosaic sequence P_LORRI (890 m/pixel), taken by the New Horizons Long-Range Reconnaissance Imager (LORRI). Mapping of cell/polygon boundaries (Fig 1c) was carried out in ArcGIS using this mosaic and additional images from P_LORRI_Stereo_Mosaic (390 m/pixel). Figs. 1a-c are simple cylindrical projections, so the scale bars are approximate. Locations of Fig 1d and Figs 2a,b are shown as insets in Fig 1a. Figure 2a is part of P_MVIC_LORRI_CA (MVIC Pan 2, 320 m/pixel), whereas Fig 2b is a segment of the LORRI portion of P_MVIC_LORRI_CA, the highest resolution image transect obtained at Pluto by New Horizons (80 m/pixel).

Stereo topography over Sputnik Planum (SP) and environs was determined using the two highest resolution Multispectral Visible Imaging Camera (MVIC) scans, P_MPan1 (495 m/pixel) and P_MVIC_LORRI_CA (MVIC Pan 2, 320 m/pixel). As MVIC is a scanning imager, each line must be individually registered carefully and pointing must be accurately known for stereo reconstruction. For Fig 1b, Pluto was assumed to be a sphere of 1187-km radius[1], and elevations were determined using an automated stereo photogrammetry method based on scene-recognition algorithms[27]. Spatial resolutions are controlled by the lower resolution MVIC scan and, using this method, are further reduced by a factor of three to five. Vertical precisions can calculated through standard stereo technique from $mr_p(\tan e_1 + \tan e_2)$, where $m$ is the accuracy of pixel matching (0.2–0.3), $r_p$ is pixel resolution, and $e_1$ and $e_2$ are the emission angles of the stereo image pair. For Fig 1b the precision is about 230 m, well suited for determining elevations of Pluto's mountains and deeper craters as well as the rim-to-floor depth of the Sputnik Planum basin. It is not sufficient to determine planum cell/polygon elevations. In the planum center the dearth of sufficient frequency topography inhibits closure of the stereo algorithm, hence the noise in the center of SP in Fig 1b.

The subtle topography of the raised cells within SP was determined from a preliminary photoclinometric (shape from shading) analysis (e.g., ref. 28), and is subject to further refinement of the photometric function for the bright cellular plains. Photoclinometry offers high-frequency topographic data at spatial scales of image resolution, but can be poorly controlled over longer wavelengths. Photoclinometry is sensitive to inherent albedo variations, but can be especially useful for investigating features with assumed symmetry, such as impact craters, which allows a measure of topographic control. The ovular domes and bounding troughs of the bright cellular plains within SP are such symmetric features, and intrinsic albedo variations are muted in the absence of dark knobs or blocks, so photoclinometry is well-suited to determining elevations across individual cells within the bright cellular plains (Figs. 1d and 2b).

**Critical Rayleigh Numbers for Convection.** Solid state viscosities $\eta$ generally follow a Arrhenius law $\eta \sim \exp(E^*/RT)$ for any given rheological mechanism, where $E^*$ is the activation energy for the deformation mechanism in question, $R$ is the gas constant, and $T$ is absolute temperature. For any given temperature and stress, one deformation mechanism generally dominates over another[29]. Critical Rayleigh number values $Ra_{cr}$ for convection, for a layer heated from below with fixed upper and lower boundary temperatures, depend on the deformation mechanism (through the power-law exponent in the stress strain-rate relation $n$) and the viscosity contrast $\Delta\eta$ across the layer due to the temperature difference $\Delta T$. In what follows we adopt an exponential viscosity law based on a linear expansion of the Arrhenius law in $E^*/RT$ (the Frank-Kamenetskii approximation) to take advantage of previous theoretical and numerical work[13,22,30,31] This is also an good approximation for the problem at hand because the temperature and viscosity contrast across a layer of volatile ices on Pluto is limited by the surface temperature of the ices on Pluto (37 K at the time of the New Horizons encounter)[1,25] and the melting temperature of $N_2$ ice (63.15 K)[12].

For an exponential viscosity law, the driving (exponential) rheological temperature scale is $\Delta T_{rh} \sim RT_i^2/E^*$, where $T_i$ a characteristic internal temperature of the convecting layer. The viscosity ratio across the layer due to temperature is then defined as $\Delta\eta =$ $\exp(\theta) = \exp(\Delta T/\Delta T_{rh})$. $Ra_{cr}$ is then approximated, for large $\theta$ and in which $T_i \approx$ the basal temperature $T_b$, by[13]

$$Ra_{cr}(n,\theta) \approx Ra_{cr}(n)\left[\frac{\exp(1)\theta}{4(n+1)}\right]^{2(n+1)/n} , \qquad (1)$$

where $Ra_{cr}(n)$ is the critical Rayleigh number for non-Newtonian viscosity with no temperature dependence (2038 for $n = 1$ and 310 for $n = 2.2$, based on numerical results for rigid upper and lower layer or sublayer boundaries[32,33]). For large $\theta$, convection occurs in the stagnant lid regime, in which convective motions are limited to a sublayer below a rigid surface. This is not the regime SP operates in, but serves as a limiting case. The transition from stagnant lid to sluggish lid convection, which does apply to SP, occurs at $\theta \approx 9$, or $\Delta\eta \approx 10^4$, for $n = 1$, and at $\theta \approx 13.8$, or $\Delta\eta \approx 10^6$, for $n = 2.2$ (ref. 13). The other convective regime limit is that of small viscosity contrast ($\Delta\eta \to 1$). For SP, with a rigid lower boundary and a free-slip upper boundary, $Ra_{cr}$ in this limit should be 1101 (ref. 34) and ~200 (estimate) for $n = 1$ and 2.2, respectively. We then estimate $Ra_{cr}(n,\theta)$ for the sluggish lid regime, following refs. 13 and 30, by linearly extrapolating in $\log\Delta\eta$-$\log Ra_b$ space between the small viscosity contrast limit and the transition to stagnant lid convection:

$$Ra_{cr}(1,\theta) \approx 1100\exp(\theta/1.78) \qquad (2b)$$

$$Ra_{cr}(2.2,\theta) \approx 200\exp(\theta/3.87) . \qquad (2b)$$

The minimum or critical volatile ice layer thickness $D_{cr}$ above which convection can occur and below which it cannot follows as[31]

$$D_{cr} \approx \left(\frac{Ra_{cr}\kappa^{1/n}\exp(E^*/nRT_i)}{3^{(n+1)/2n}A^{1/n}\rho g\alpha\Delta T}\right)^{n/(n+2)} , \qquad (3)$$

where $\kappa$, $\rho$, and $\alpha$ are, respectively, the thermal diffusivity, density, and volume thermal expansion coefficient of the ice, and $A$ is the preexponential coefficient in the stress strain-rate relationship. For $N_2$ ice, this is either measured directly[4] or estimated theoretically[12]. The numerical factor in the denominator comes from the definition of viscosity and the conversion from laboratory geometry ($A$ is measured in uniaxial compression) to the generalized flow law. For sluggish lid convection, we approximate $T_i$ as $T_b - \Delta T/2$, which is a slight underestimate for the problem under discussion, but one that makes $D_{cr}$ in equation (3) an upper bound on the minimum thickness for convection.

Equation (3) does not explicitly depend on ice grain size $d$. The power-law exponent reported for nitrogen ice deformation ($n \approx 2.2$)[4] suggests a grain-size sensitive regime such as a grain boundary sliding, as opposed to a purely dislocation creep or climb mechanism (which would be grain-size independent)[35]. Grain sizes in the nitrogen ice deformation experiments were not reported[4], but it was noted that the grain sizes of similar experiments on methane ice were a few mm. This is a not atypical grain size for convecting upper mantle rock, deep polar glacial ice on Earth, and is plausible for convecting water ice within icy satellites of the outer Solar System[36], so without further information we utilize the deformation experiment results for nitrogen[4] as is. Notably, however, in order for $N_2$ ice to be identified spectroscopically at all on Pluto, very long optical path lengths are required (>> 1 cm)[37], so the grain sizes of the convecting ice within SP may be much larger than a few millimeters. Because grain-size-sensitive rheologies typically have viscosities proportional to $d^2$ or $d^3$, the presumed $N_2$ ice in SP may be much more viscous than in the reported experiments[4]. On the other hand, the presence of convective cells in SP implies that the viscosity is not arbitrarily large. Grain sizes in the annealed, convecting ice are likely determined by stress levels and the presence of contaminants (such as bits of water ice or tholins) and minor phases (such as $CH_4$-rich ice)[36]. Diffusion creep is also grain-size dependent, and in evaluating $N_2$ diffusion creep for comparison with Fig. 3 we adopt $d$ = 1 mm as a nominal value, noting that for volume diffusion $D_{cr}$ scales as $d^{2/3}$. The minimum thickness for convection by volume diffusion would plot off the graph in Fig 3 to the upper right for $d$ = 1 mm. Only if $d$ were much smaller would $D_{cr}$ for volume diffusion be comparable to that shown in Fig 3.

Regarding the potential role of CO ice in SP, we note the near-perfect solid solution between solid $N_2$ and CO, and close similarities in density, melting temperature and electronic structure[15]. Hence, if the deeper ice in SP were actually dominantly CO, it would behave much the same as pure $N_2$ ice, with the proviso that an $N_2$-CO ice solid solution under Pluto conditions would, for CO fractions greater than 10%, crystallize in the ordered $\alpha$ phase, as opposed to the disordered $\beta$ phase of $N_2$. We expect $\alpha$-phase CO to be stiffer than its $\beta$-phase counterpart, based on the viscosity differences between ordered and disordered water ice phases[38]. We stress, however, that the surface of SP, whatever its precise composition, is itself not in the $\alpha$ phase, for if so the 2.16-$\mu$m $N_2$ absorption feature would not be observed[37].

Regarding the potential role of $CH_4$ ice in SP, deformation experiments indicate similar behavior to that of $N_2$ ice, but $CH_4$ ice appears to be about 25 times more viscous than $N_2$ ice (i.e., $A$ is ~25 times larger at the same $T$ and differential stress)[4], and with a similar power-law index $n$. The minimum or critical $D_{cr}$ for convection within SP from equation (3) would than be about double that in Fig. 3 if SP were in fact filled with $CH_4$ ice, so the convection hypothesis is just as valid for $CH_4$ ice as for $N_2$ ice. The geological and compositional data point to an $N_2$-dominated layer, however, as discussed in the main text.

Applying rheological data obtained in laboratory conditions to geological problems often requires extrapolation to different stress and strain conditions. For convection these conditions are lower stresses and strain rates. This is true whether one is modeling convection in the mantle of the Earth or another terrestrial planet (with peridotite), in the icy satellites of the giant planets (with water ice), or in the present case of Sputnik Planum (with volatile ices such as $N_2$). The extrapolation is valid if the same stress mechanism or mechanisms dominate at the extrapolated conditions[38,39]. The $n$ values reported for laboratory deformation of $N_2$ ice and $CH_4$ ice[4] are low enough (2.2 ± 0.2 and 1.8 ± 0.2, respectively) that it seems implausible that some power-law, dislocation mechanism ($n$ ~ 3-5) becomes dominant at lower stresses. Rather, the only likely transition would be, depending on $T$, to volume or grain-boundary diffusion ($n$ =1), which we already consider. Regardless, our understanding of $N_2$ and other volatile ice rheology could be greatly improved, especially any dependence on grain size.

Solid $N_2$ material parameters for Fig. 2 are as follows: $\kappa = 1.33 \times 10^{-7}$ m$^2$ s$^{-1}$, $\alpha = 2 \times 10^{-3}$ K$^{-1}$, $E^* = 3.5$ kJ mole$^{-1}$ ($n = 2.2$), $E^* = 8.6$ kJ mole$^{-1}$ ($n = 1$), $A = 3.73 \times 10^{-12}$ Pa$^{-2.2}$ s$^{-1}$ ($n = 2.2$), $A = 1.52 \times 10^{-7} \times (d/1\ \text{mm})^{-2} \times (T/50\ \text{K})^{-1}$ Pa$^{-1}$ s$^{-1}$ ($n = 1$), $\rho = 1000 - 2.14(T - 36\ \text{K})$ kg m$^{-3}$, and for the heat flow calculations, conductivity $k = 0.2$ W m$^{-1}$ K$^{-1}$ (refs. 4,15,21). Pluto's surface gravity is 0.617 m s$^{-2}$ (ref. 1).

**Convection Simulations.** Numerical convection calculations were carried out with the well-benchmarked fluid dynamics finite element code CitCom[22]. CitCom solves the equations of thermal convection of an incompressible fluid in the Boussinesq approximation and at infinite Prandtl number. CitCom can solve the thermal convection equations using an Arrhenius viscosity or an exponential law (the Frank-Kamenetskii approximation). We utilized this latter approximation here, for both Newtonian (stress-independent) and non-Newtonian viscosities, to best compare our results with those in the literature[5,13,22,30].

We first simulated solid state convection with a Rayleigh number $Ra = 2 \times 10^4$ but with a non-temperature-dependent viscosity, in a very wide, rectangular 32 x 1 domain, with 2048 × 64 elements, to allow natural selection of convection cell aspect ratios (widths of convective cells divided by layer depth). Temperatures at the top and bottom of the domain were fixed. Free slip was assumed at the surface, no slip at the base (the volatile ice layer is in contact with a rigid, water-ice basement), and periodic, free-slip boundary conditions along the sides of the domain. Velocities normal to domain edges in all cases were zero. Simulations were allowed to reach steady state. Calculations were carried out for Newtonian, isoviscous flow, and for non-Newtonian ($n = 2.2$) flow, both with the same Rayleigh number. In both cases the planforms were characteristic of their entire respective domains, and the aspect ratios for the convective cells for both simulations were close to 1, as expected from theory and previous results. (For example, the critical wavelength at $Ra = Ra_{cr}$ for a plane layer heated from below, with boundary conditions appropriate for convection within SP, is 2.34 times the layer depth[34].)

A suite of calculations was then carried at a variety of $Ra_b$ and top-to-bottom viscosity ratios $\Delta\eta = \exp(\theta) = \exp(E^*\Delta T/RT_b^2)$, where $Ra_b$ is defined as the basal $Ra$ (i.e., $T$ in equation (1) of the main text = $T_b$). Rectangular 12 x 1 domains, with 768 × 64 elements, were utilized, with the same boundary conditions as above. A smaller number of calculations were also run with a free-slip lower boundary, for benchmarking with examples presented in ref. 5, and to simulate convection where the SP ice is at or near melting at its base. All runs in this suite were Newtonian, and while convective aspect ratios were not predictable from theory alone, they were expected to be much greater than 1 (ref. 5). In all cases simulations were allowed to reach steady state, or if time-dependent, to reach characteristic state behavior.

Our present survey covers a range of $Ra_b$ between $10^4$ and $10^6$, and a range in $\Delta\eta$ between 150 and 3000. This reflects our judgment that the convective regime represented by the cells in SP ranges from the obviously convectively unstable to the subcritical (i.e., stable) at the periphery of the basin (e.g., Fig. 2b). The transition from cellular to non-cellular plains could reflect several things, including shallowing of the volatile ice layer, lower heat flow, and in the case of non-Newtonian flow, an insufficient initial temperature perturbation[13,31,33]. The simplest explanation, however, for smaller cell sizes with distance from the center of SP (Fig. 1c), and then a transition to level plains (no cells) towards the south (e.g., Fig. 2b), is that the SP basin is shallower towards its margins, and particularly shallow towards its southern margin. This is consistent with the expected basin topography created by an oblique impact to the SSW[40]. The less well defined cellular structure in the very center of SP may, in contrast, reflect the deeper center of the basin, implying a larger $Ra$ for the $N_2$ ice layer there and more chaotic, time dependent convection.

Our numerical simulations are carried out in in terms of dimensionless parameters, and do not presuppose any particular values for the depth of the SP volatile ice layer or Pluto's heat flow, etc. They can be dimensionalized to determine if various measureable or estimable quantities are matched or are at least self-consistent. Depths and lengths scale as $D$, velocities as $\kappa/D$, stresses as $\eta_b\kappa/D^2$ ($\eta_b$ is the basal viscosity), and heat flows as $k\Delta T/D$ (ref. 22). For example, for a given simulation, $D$ can be scaled from surface cell size. Then different heat flows imply different $\Delta T$. At fixed $D$ and $Ra_b$, $\eta_b$, stresses, and dynamic topography all scale with $\Delta T$.

**Code Availability.** CitCom is freely available, in the version CitComS, released under a General Public License and downloadable from the Computational Infrastructure for Geodynamics (geodynamics.org).

**Data Availability**. All spacecraft data and higher-order products presented in this paper will be delivered to NASA's Planetary Data System (https://pds.nasa.gov) in accordance with the schedule established by NASA and the New Horizons project. This will occur in a series of stages in 2016 and 2017 because of the time required to fully downlink and calibrate the data set.

**Figure Captions.**

**Figure 1 | Image, Topographic, and Maps Views of Sputnik Planum, Pluto. a)** Base map; **b)** Stereo-derived topography, showing that Sputnik Planum (SP) lies within a kilometers deep basin. SW-NE banding and central basin "speckle" are artifacts or noise (Methods); elevations are relative; **c)** Map of troughs (black lines), which define cell boundaries (note enlarged scale compared with **a** and **b**). Cell size increases and/or becomes less well connected towards SP center, consistent with a thickened $N_2$ ice layer there. Aquamarine shading indicates "bright cellular plains," within which troughs are topographically defined; **d)** 350 m/pixel MVIC inset (see **a)** that shows cellular/polygonal detail.

**Figure 2 | High Resolution Images of Cellular Terrain within Sputnik Planum, Pluto. a)** Kilometer-scale hills appear to emanate from uplands to the east, and are likely darker water-ice blocks and methane-rich debris that have broken away and are being carried by denser, $N_2$-ice-dominated glaciers into SP, where they become subject to the convective motions of SP ice, and are pushed to the downwelling edges of the cells. **b)** Part of the highest resolution image sequence taken by New Horizons (80 m/px), surface texture (e.g., pitting) concentrates towards cell boundaries and in regions apparently unaffected by convection (such as at right, see text).

**Figure 3 | Minimum thickness for convection in a layer of solid $N_2$ ice on Pluto, as a function of basal temperature.** Convection can occur above the red curve provided a sufficient perturbation exists. Limit is based on numerical and laboratory experiments and theory and creep measurements for nitrogen ice (Methods). Basal temperatures due to conductive heat flow (3 mW $m^{-2}$) from Pluto are shown for comparison. For ~present-day chondritic heat flows, basal temperatures exceed the convective threshold for layer thicknesses ≳500 meters. In contrast, the minimum thickness for convection by volume diffusion creep would plot off the graph to the upper right.

**Figure 4 | Example numerical model of $N_2$ ice convection in Sputnik Planum. a)** Temperature field showing large-aspect-ratio plumes and downwellings. Basal Rayleigh number $Ra_b$ and viscosity ratio $\Delta\eta$ are indicated. White contour denotes the median temperature; **b-d)** Corresponding horizontal surface velocities, vertical normal stresses and dynamic topography, and surface heat flows. Non-dimensional values are shown are shown at left, and dimensional values at right assuming $D$ = 4.5 km and $\Delta T$ = 20 K. Calculated topography matches the scale seen within the bright cellular plains, and the average heat flow is consistent with radiogenic heat production in Pluto’s rock fraction.

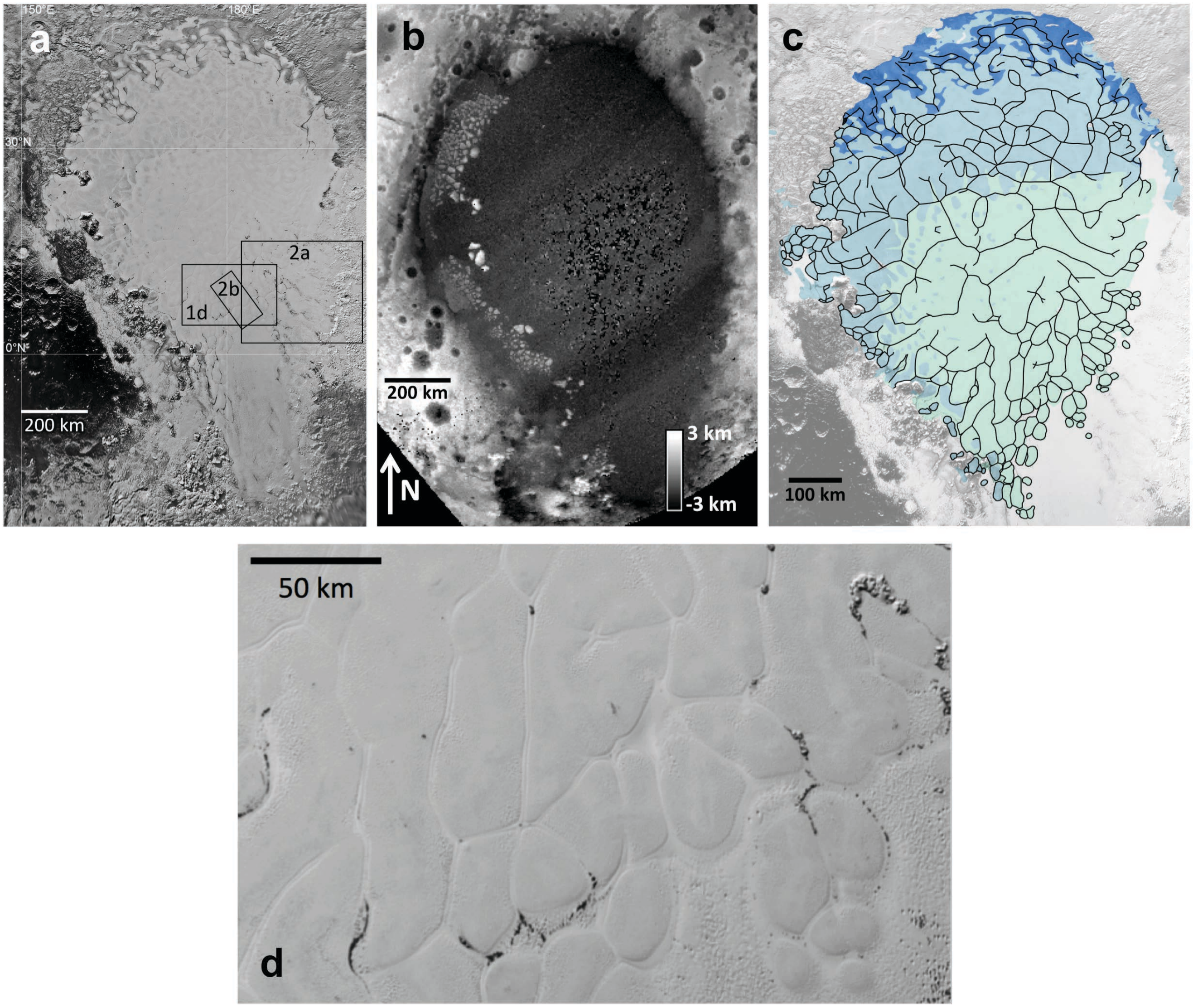

**Figure 1**

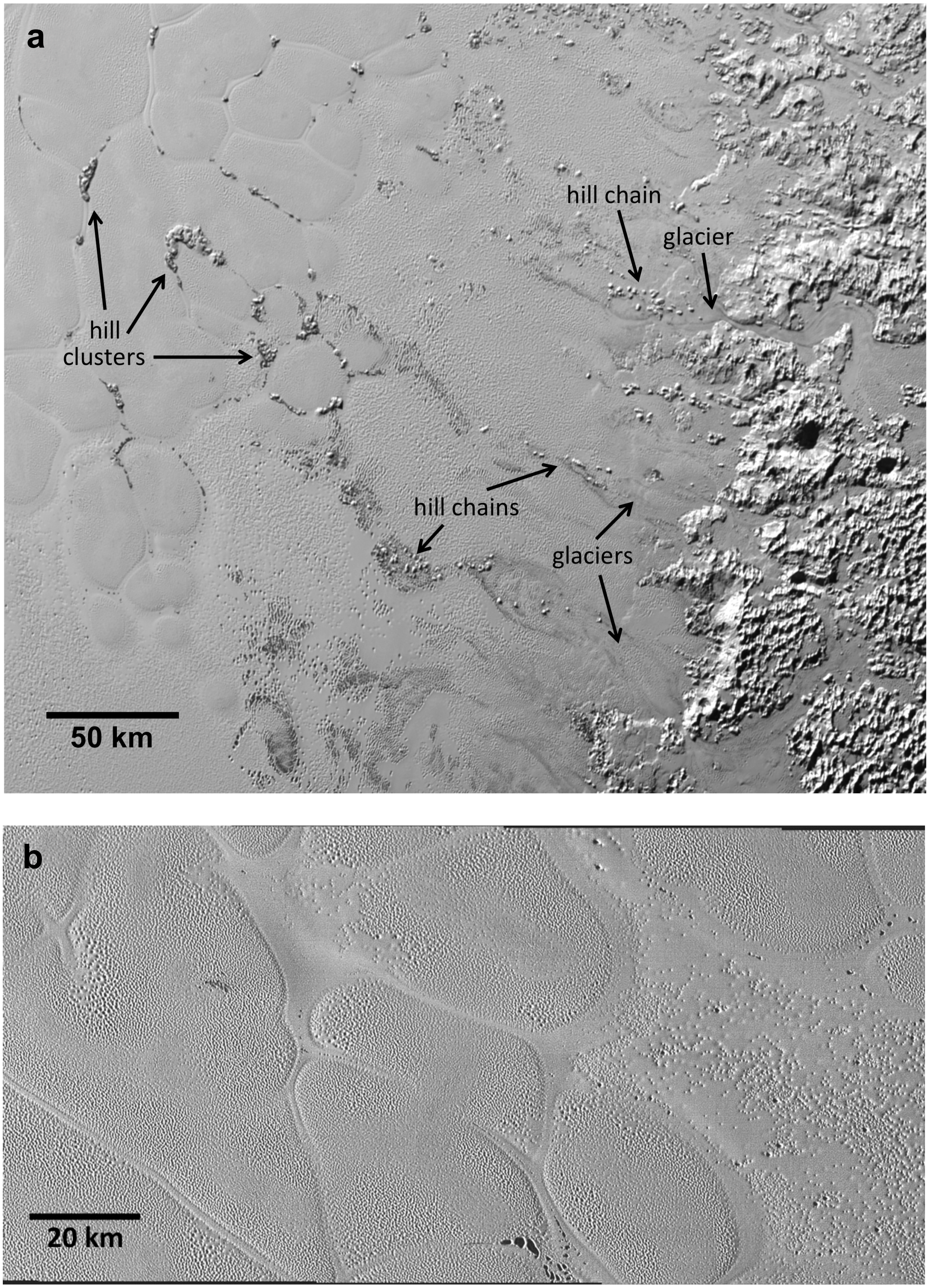
a
hill clusters
hill chain
glacier
hill chains
glaciers
50 km
b
20 km

**Figure 2**

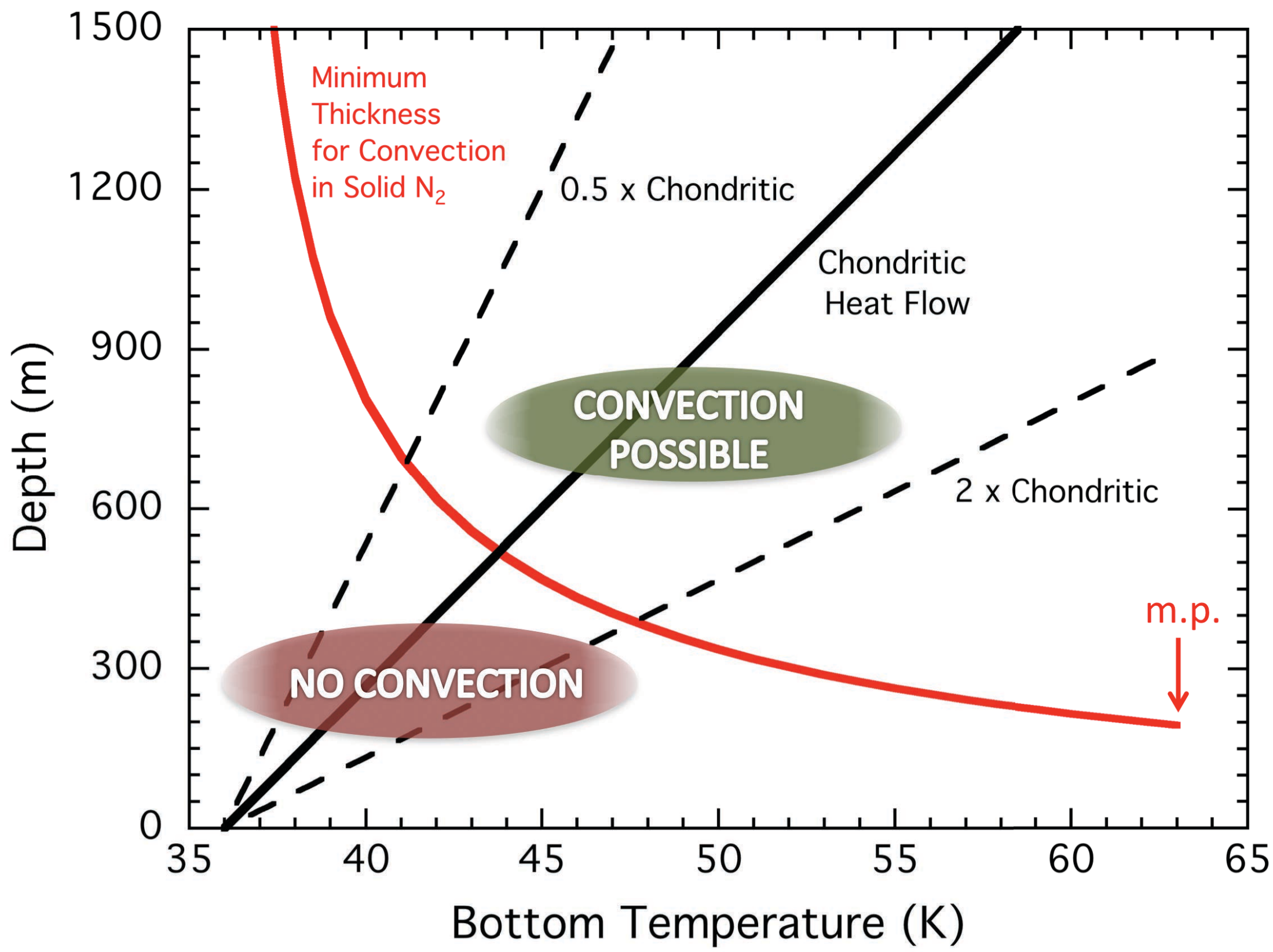
1500
1200
900
600
300
0
35
40
45
50
55
60
65
Depth (m)
Bottom Temperature (K)
Minimum
Thickness
for Convection
in Solid N2
0.5 x Chondritic
Chondritic
Heat Flow
CONVECTION
POSSIBLE
2 x Chondritic
m.p.
NO CONVECTION

**Figure 3**

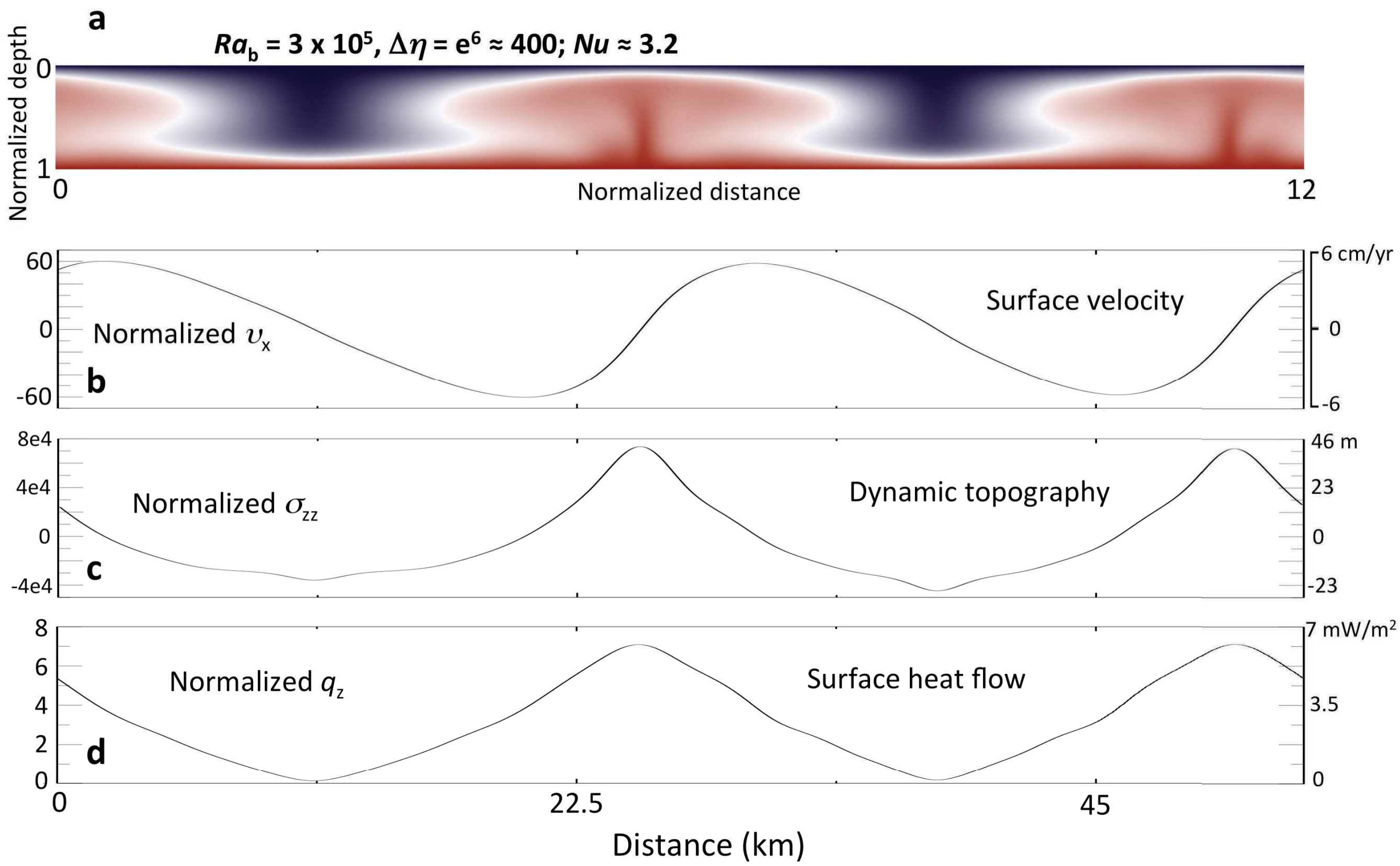
a
$Ra_b$ = 3 x 10$^5$, $\Delta\eta$ = e$^6$ ≈ 400; $Nu$ ≈ 3.2
Normalized depth
0
1
0
Normalized distance
12
60
0
-60
Normalized $\upsilon_x$
b
Surface velocity
6 cm/yr
0
-6
8e4
4e4
0
-4e4
Normalized $\sigma_{zz}$
c
Dynamic topography
46 m
23
0
-23
8
6
4
2
0
Normalized $q_z$
d
Surface heat flow
7 mW/m$^2$
3.5
0
0
22.5
45
Distance (km)

**Figure 4**

**New Horizons Geology and Geophysics Imaging Theme Team members:**

J.M. Moore[1], W.B. McKinnon[2], J.R. Spencer[3], R. Beyer[1], R.P. Binzel[25], M. Buie[3], B. Buratti[4], A. Cheng[5], D. Cruikshank[1], C.Dalle Ore[1], A. Earle[25], R. Gladstone[6], W. Grundy[7], A.D. Howard[8], T.Lauer[9], I. Linscott[10], F. Nimmo[11], C. Olkin[3], J. Parker[3], S. Porter[3], H. Reitsema[12], D. Reuter[13], J.H. Roberts[5], S. Robbins[3], P.M. Schenk[14], M. Showalter[15], K. Singer[3], D. Strobel[16], M. Summers[17], L. Tyler[10], H. Weaver[5], O.L. White[1], O.M. Umurhan[1], M. Banks[18], O. Barnouin[5], V. Bray[19], B. Carcich[20], A. Chaikin[21], C. Chavez[1], C. Conrad[3], D. Hamilton[22], C. Howett[3], J. Hofgartner[20], J. Kammer[3], C. Lisse[5], A. Marcotte[5], A. Parker[3], K. Retherford[6], M. Saina[5], K. Runyon[4], E. Schindhelm[3], J. Stansberry[23], A. Steffl[3], T. Stryk[24],. H. Throop[3], C. Tsang[3], A. Verbiscer[8], H. Winters[5], A. Zangari[3], S.A. Stern[3], H.A. Weaver[5], C.B. Olkin[3], L.A. Young[3], K.E. Smith[1]

[1]National Aeronautics and Space Administration (NASA) Ames Research Center, Moffett Field, California 94035, USA. [2]Department of Earth and Planetary Sciences and McDonnell Center for the Space Sciences, Washington University in St Louis, Saint Louis, Missouri 63130, USA. [3]Southwest Research Institute, Boulder, Colorado 80302, USA. [4]NASA Jet Propulsion Laboratory, Pasadena, California 91019, USA. [5]Johns Hopkins University Applied Physics Laboratory, Laurel, Maryland 20723, USA. [6]Southwest Research Institute, San Antonio, Texas 78238, USA. [7]Lowell Observatory, Flagstaff, Arizona 86001, USA. [8]University of Virginia, Charlottesville, Virginia 22904, USA. [9]National Optical Astronomy Observatory, Tucson, Arizona 85719, USA. [10]Stanford University, Stanford, California 94305, USA. [11]Department of Earth and Planetary Sciences, University of California Santa Cruz, Santa Cruz, California 95064, USA. [12]B612 Foundation, Mill Valley, California 94941, USA. [13]NASA Goddard Space Flight Center, Greenbelt, Maryland 20771, USA. [14]Lunar and Planetary Institute, Houston, Texas 77058, USA. [15]The SETI Institute, Mountain View, California 94043, USA. [16]The Johns Hopkins University, Baltimore, Maryland 21218, USA. [17]George Mason University, Fairfax, Virginia 22030, USA. [18]Planetary Science Institute, Tucson, Arizona 85719, USA. [19]University of Arizona, Tucson, Arizona 85721, USA. [20]Cornell University, Ithaca, New York 14853, USA. [21]Arlington, Vermont 05250, USA. [22]University of Maryland, College Park, Maryland 20742, USA. [23]Space Telescope Science Institute, Baltimore, Maryland 21218, USA. [24]Roane State Community College, Oak Ridge, Tennessee 37830, USA. [25]Massachusetts Institute of Technology, Cambridge, Massachusetts 02139, USA.